\documentclass[12pt]{iopart}

\usepackage[dvipdfmx]{graphicx}% Include figure files
\usepackage{color}
\usepackage{cite}

\begin{document}

\title["Quest for the non-perturbative magnetic field effects in 1000-Tesla"]
{Magnetic-field-induced insulating behavior in black phosphorus under pressure} %Title of the paper

\author{Kazuto Akiba$^{1*}$, Yuzuki Sega$^1$, Yuichi Akahama$^2$, Yuta Seo$^3$, Tomoki Machida$^3$, and Masashi Tokunaga$^4$}

\address{$^1$Faculty of Science and Engineering, Iwate University, Morioka, Iwate 020-8551 Japan.\\
$^2$Department of Material Science, Graduate School of Science, University of Hyogo, Kamigori, Hyogo 678-1297, Japan.\\
$^3$Institute of Industrial Science (IIS), The University of Tokyo, Meguro, Tokyo 153-8505, Japan.\\
$^4$The Institute for Solid State Physics (ISSP), The University of Tokyo, Kashiwa, Chiba, 277-8581, Japan.
}

\vspace{10pt}
\begin{indented}
\item[]$^*$ E-mail : akb@iwate-u.ac.jp
\end{indented}

\begin{abstract}
We investigated the out-of-plane magnetoresistance of pressurized black phosphorus (BP) with a longitudinal field configuration.
Despite the absence of the Lorentz force in the present configuration,
we observed a significant enhancement of magnetoresistance marked with a clear onset field
in both the semiconducting (1.1 GPa) and semimetallic (1.3 GPa) phases.
The insulating behavior observed
near the semiconductor-semimetal transitio pressure is possibly associated with emergence of an excitonic phase,
which has been suggested in a recent theoretical study.
BP under finely tuned pressure can be a candidate to realize
the field-induced electronic phase transition in a moderate magnetic field below 9 T. 
\\
\end{abstract}

\section{Introduction}
Exotic electronic phases in quantum limit state have been one of the central issues in high-field studies.
In the quantum limit state, where only the lowest Landau level provides the charge carrier,
the cyclotron motion is confined to the smallest orbit.
Due to the small kinetic energy, the correlation effect between charged carriers plays a primary role in the quantum limit state.
Moreover, the system can be unstable against the 2$k_F$ instability
for the quasi one-dimensional Landau subband structure in high fields \cite{Yoshioka_1981}.
Thus, semimetals and narrow gap semiconductors with small carrier density provide
a playground to explore novel electronic phases,
e.g.,
excitonic insulator phase driven by strong electron-hole correlations
and
density wave phase associated with a nesting of the Fermi surface.
Realization of these phases results in an insulating behavior characterized by an abrupt increase of resistivity in high fields.
A typical example is graphite,
which exhibits successive insulating phases in the magnetic field ($B$) parallel to the $c$-axis \cite{Zhu_2019}.
Intriguingly, similar field-induced insulating behavior has recently been reported in the BiSb alloy \cite{Kinoshita_2023,Yamaguchi_2025}.
Thus, elucidating the universal aspect of the electronic state realized in the quantum limit state is of primary importance.

\begin{figure}[]
\centering
\includegraphics[clip,width=0.65\columnwidth]{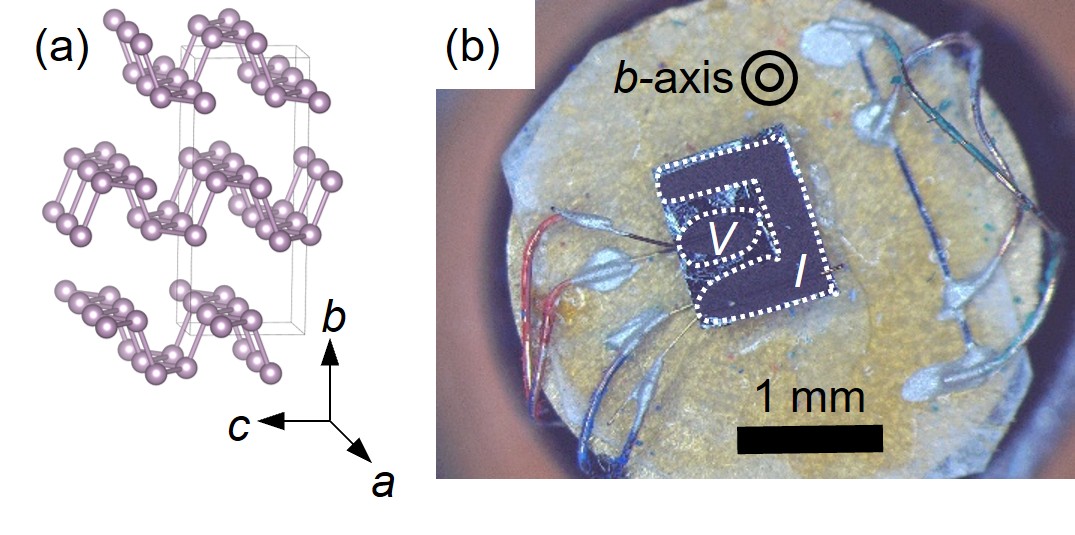}
\caption{(a) Crystal structure of BP with crystal axes.
(b) Typical setup of out-of-plane resistivity measurement using the four-probe method under pressure.
$I$ and $V$ represent the current and voltage terminals attached to one side of a sample.
\label{f1}}
\end{figure}

An alternative platform to study the exotic electronic states is black phosphorus (BP) under pressure.
The crystal structure of BP is shown in Fig. \ref{f1}(a), in which the puckered honeycomb layers stack along the $b$-axis.
the zigzag- and armchair-direction of the puckered honeycomb lattice is represented as $a$- and $c$-axis, respectively.
BP is a typical narrow-gap semiconductor with the energy gap of $\sim335$ meV at ambient pressure \cite{Akahama_1983},
and it turns into a nodal ring semimetal (NRS) under pressure \cite{Akiba_2024}.
The tunability of carrier density in BP provides an opportunity to realize the exotic electronic
states in moderate magnetic fields.

A previous study reported a sudden increase in out-of-plane resistivity at $\sim 17$ T
in the semimetallic phase with a longitudinal field configuration \cite{Sun_2018},
i.e., both $B$ and electric current $I$ are parallel to the $b$-axis ($B \parallel I \parallel b$).
The anomaly shows significant temperature dependence, which is similar to the cases of graphite and BiSb.
In contrast, the in-plane resistivity
($B \parallel b$ and $I$ within the $ac$ plane)
shows an unobvious peak, which can be discernible in the second derivative.
Although this field-induced insulating behavior was interpreted
as a formation of a density-wave phase due to 2$k_F$ instability,
the origin remains unclear.

In the case of BP, the out-of-plane transport has hardly been studied besides Ref. \cite{Sun_2018},
in contrast to the relatively abundant knowledge on the in-plane transport.
Considering the striking difference between in-plane and out-of-plane magnetotransport mentioned above,
a comprehensive investigation of the out-of-plane magnetotransport under pressure is important
not to overlook the clue of field-induced electronic phase transition in BP.
Hence, we investigated the out-of-plane magnetotransport of BP using finely tuned pressure up to 1.5 GPa.

\section{Experimental methods}
Single crystals of BP were synthesized under high pressure \cite{Endo_1982,Akahama_2020}.
Typical setup to measure the out-of-plane resistivity ($\rho_b$)
using four-probe method is shown in Fig. \ref{f1}(b),
which was intended to prevent inhomogeneity of the current flow
and obtain reliable resistivity.
High pressure up to 1.5 GPa was applied using a piston-cylinder-type pressure cell.
Daphne7373 oil was utilized as pressure-transmitting medium,
and the pressure of the sample space was determined
by the superconducting transition temperature of lead \cite{Eiling_1981}.
Magnetic fields were generated using superconducting magnets.
Low temperature was realized using a sorption pump-type refrigerator
or a variable temperature insert of the magnet.

\section{Results and discussion} 

\begin{figure}[]
\centering
\includegraphics[clip,width=0.9\columnwidth]{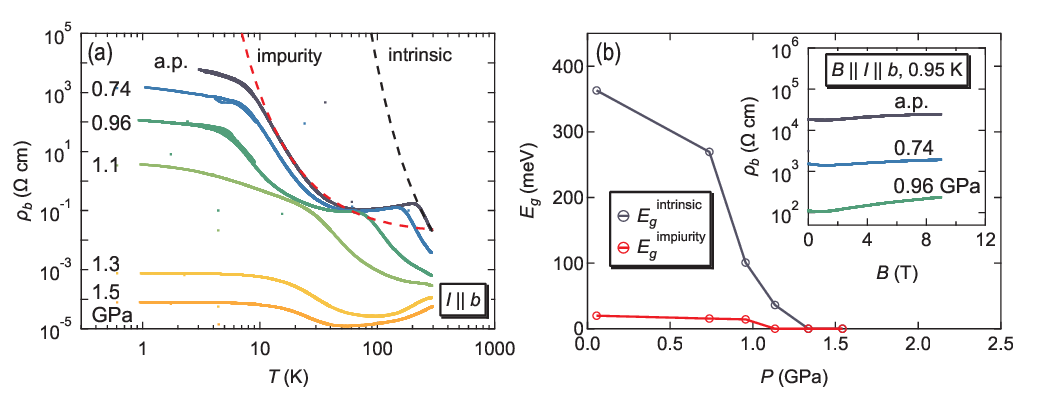}
\caption{(a) Temperature dependence of $\rho_b$ under pressures up to 1.5 GPa.
The black and red broken curves represent the fitting curves assuming the Arrhenius model.
(b) Pressure dependence of the intrinsic ($E_g^{\mathrm{intrinsic}}$) and impurity ($E_g^{\mathrm{impurity}}$) gaps
deduced by Arrhenius fitting.
The inset shows the field dependence of $\rho_b$ at 0.95 K measured sufficiently below the semiconductor-semimetal transition pressure.
\label{f2}}
\end{figure}

Firstly, we show in Fig. \ref{f2}(a) the temperature ($T$) dependence of $\rho_b$ at various pressures.
As shown by the broken lines,
Arrhenius-type thermal excitations across the intrinsic band gap (black)
and between the valence band and impurity level (red) were clearly identified.
The pressure dependence of the intrinsic band gap ($E_g^{\mathrm{intrinsic}}$)
and impurity gap ($E_g^{\mathrm{impurity}}$)
are deduced by the fitting procedures on the resistivity assuming $\rho_b\propto \exp[E_g/(2k_B T)]$,
where $k_B$ is the Boltzmann constant.
The result is shown in Fig. \ref{f2}(b).
$E_g^{\mathrm{intrinsic}}$ of $\sim 360$ meV at ambient pressure is consistent with the previous study \cite{Akahama_1983}.
Both $E_g^{\mathrm{intrinsic}}$ and $E_g^{\mathrm{impurity}}$ monotonically decrease as a function of pressure.
Above 1.3 GPa,
the temperature dependence of the resistivity turns to be metallic
in 100 K $<T<$300 K, indicating the collapsing of $E_g^{\mathrm{intrinsic}}$.
Besides, $\rho_b$ at low temperature significantly decreased by more than three orders of magnitude
between 1.1 and 1.3 GPa.
The above results indicate that the semiconductor-semimetal (SC-SM) transition
occurs between 1.1 and 1.3 GPa.
It is also worth mentioning that
$\rho_b$ becomes comparable to in-plane resistivity in the semimetallic phase (10$^{-3}$-10$^{-5}$ $\Omega$ cm) \cite{Akiba_2024},
despite the anisotropic layered crystal structure.

The inset of Fig. \ref{f2}(b) shows
the $B$-dependence of $\rho_b$ at 0.95 K measured sufficiently below the SC-SM transition pressure.
The magnetoresistance (MR) effect, 
defined by the ratio of $\rho_b$ between 9 T and 0 T was 1.3, 1.3, and 2.2 at ambient pressure, 0.74 GPa, and 0.96 GPa, respectively.
Such a weak MR is common to the longitudinal configuration,
where the Lorentz force does not work within a sense of classical framework. 

\begin{figure}[]
\centering
\includegraphics[clip,width=0.9\columnwidth]{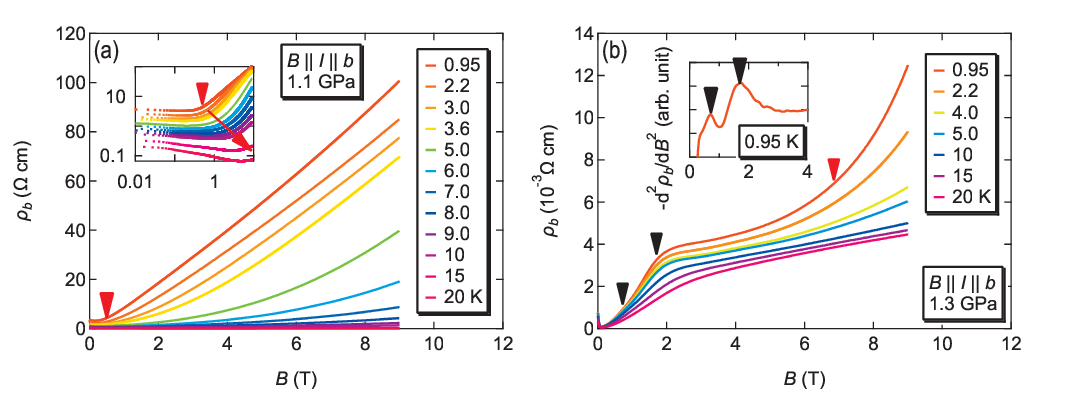}
\caption{(a) Field dependence of $\rho_b$ at 1.1 GPa.
The inset shows the log-log plot of $\rho_b$.
The red arrow indicates the onset field of the resistivity increase.
(b) Field dependence of $\rho_b$ at 1.3 GPa.
The inset shows the second derivative of $\rho_b$ at 0.95 K.
The black arrows represent the resistivity peaks caused by SdH oscillation,
and the red arrow indicates the onset of field-induced insulating behavior.
\label{f3}}
\end{figure}

Then, we focus on the MR just below the SC-SM transition.
Figure \ref{f3}(a) shows the $B$-dependence of $\rho_b$ at 1.1 GPa.
In contrast with the former case,
we observed a significantly large MR effect at this pressure, reaching 28 at 9 T and 0.95 K.
At the onset field shown in the red arrow, the MR turns to increase,
and continues to grow without any tendency towards saturation.
As shown in the log-log plot in the inset of Fig. \ref{f3}(a),
the onset field shows noticeable temperature dependence,
which systematically increases as the temperature increases.
This suggests an electronic phase transition driven by many-body effect.
We also confirmed that the above behavior can be reproduced in another piece of sample with a different setup.
We can see clear semiconducting temperature dependence in Fig. \ref{f2}(a)
and no SdH oscillation in Fig. \ref{f3}(a).
Thus, the Fermi surface does not exist at this pressure.
We note that above insulating behavior emerges within the semiconducting phase,
which is in a different situation from the previous study \cite{Sun_2018} focusing on the semimetallic phase.

Next, we show the MR just above the SC-SM transition pressure.
Figure \ref{f3}(b) shows
the $B$-dependence of $\rho_b$ at 1.3 GPa.
We can recognize a hump-like structure at 1.7 T.
As clearly seen in the second derivative of the resistivity in the inset of Fig. \ref{f3}(b),
this is a part of the SdH oscillation with quite a small frequency of $F\sim$ 0.84 T.
Thus, the system has a Fermi surface and enters into the quantum limit state above 1.7 T.

Besides the SdH oscillation below 2 T,
we observed a sudden upturn of the resistivity at $\sim$ 7 T.
The enhancement of the resistivity is strongly suppressed by raising the temperature,
which is no longer discernible above 10 K.
This is in contrast with the SdH peak at 1.7 T,
which is still clearly visible at 10 K. 
Since the quantum limit state is attained above 1.7 T,
the anomaly at 7 T cannot be a part of SdH oscillation.
The field-induced insulating behavior with strong temperature dependence
resembles the anomaly
seen at 17 T in a previous study \cite{Sun_2018}.
Particularly noteworthy is the considerable reduction of the onset field in the present study,
which can be realized below 9 T.
The possible reason for this moderate onset field is
smaller carrier density represented by the SdH frequency of $F\sim$ 0.84 T,
compared to the previous study of $F\sim 3.6$ T \cite{Sun_2018}.

Finally, we would like to consider the origin of the observed insulating behavior
based on the recent theoretical work \cite{Wang_2024}.
Wang \textit{et al.} investigated the electronic instability of BP realized under high pressure and high magnetic field.
The situation assumed in this theoretical work is a NRS phase,
whose Fermi surface has been certified by our recent field-angular-resolved SdH oscillation measurement \cite{Akiba_2024}.
Thus, we can reasonably interpret the results obtained just after the metallization
[Fig. \ref{f3}(b)] based on this theoretical work.
According to the predicted phase diagram in hole-doping density vs. magnetic field space,
the ground state in semimetallic BP under high fields depends on the hole-doping amount.
In high doping case (more than $3.5\times 10^{17}$ cm$^{-3}$),
a charge density wave phase can be realized
just after the system enters into the quantum limit state.
In low doping case (less than $5\times 10^{16}$ cm$^{-3}$), on the other hand,
an excitonic insulator phase is expected
above the field-induced NRS-semiconductor reentrant transition.
From the observed SdH frequency of $F\sim$ 0.84 T,
we can estimate that the carrier density is an order of $10^{15}$ cm$^{-3}$ assuming a spherical Fermi surface.
Thus, the present situation corresponds to the low hole-doping density,
and the realization of an excitonic insulator phase is the most probable scenario.

With regard to the large nonsaturating longitudinal magnetoresistance observed at 1.1 GPa [Fig. \ref{f3}(a)],
this situation lies beyond the prerequisite condition assumed in the theoretical work,
since no Fermi surface exists at 1.1 GPa.
Nevertheless, we speculate that this behavior provides a clue to the emergence of an excitonic insulator in the semiconducting phase.
At 1.1 GPa, $E_g^{\mathrm{intrinsic}}$ is exceedingly small.
According to the Landau subband calculation \cite{Wang_2024}, the gap is assumed to increase as $B$ increases.
This enlargement of the gap leads to a reduction in the number of carriers, i.e., a weakening of the screening effect, which is favorable for exciton formation.
Concrete mechanism of the anomalous behavior observed near the SC-SM transition pressure
should be further clarified by subsequent future work.

\section{Conclusion}
We investigated the out-of-plane transport of black phosphorus under finely tuned pressure up to 1.5 GPa.
We identified the semiconductor-semimetal transition pressure of the investigated sample located between 1.1 GPa and 1.3 GPa.
At pressures sufficiently lower than the critical pressure, the longitudinal magnetoresistance effect was quite weak,
which is consistent with the absence of the Lorentz force.
In the semiconducting phase with an almost collapsing band gap, on the other hand,
we observed a large nonsaturating magnetoresistance effect, whose onset field shows significant temperature dependence.
Just after the metallization showing quite a small Shubnikov-de Haas frequency of $\sim$ 0.84 T,
an abrupt upturn of the resistivity was observed at 7 T, which is significantly suppressed by raising the temperature from 0.95 K to 10 K. 
The insulating behavior observed near the semiconductor-semimetal transition pressure
might be related to a realization of an excitonic insulator phase driven by strong electron-hole correlation,
which has been suggested in a recent theoretical study.
We demonstrated that pressurized black phosphorus can be a promising platform to study
unconventional electronic phase transition in easily accessible magnetic fields lower than 9 T.

\section*{Acknowledgments}
We thank H. Sakai and Y. Fuseya for valuable discussion and comments.
This study was supported by JSPS KAKENHI Grant No. 23H04862.

\end{document}